\documentclass[aip,pop,reprint,amsmath,amssymb]{revtex4-2}

\usepackage{graphicx}
\usepackage{bm}
\usepackage{xcolor}
\usepackage{microtype}
\usepackage{capt-of}
\newcommand{\newrev}[1]{#1}
\newcommand{\termrev}[1]{#1}
\newcommand{\concrev}[1]{#1}
\newcommand{\ackrev}[1]{#1}

\begin{document}

\title{Polarization-resolved attosecond gamma-ray emission from few-cycle laser interactions with cone targets}

\author{De-Sheng Zhang}
\thanks{These authors contributed equally to this work.}
\affiliation{Key Laboratory of Beam Technology of the Ministry of Education, and School of Physics and Astronomy, Beijing Normal University, Beijing 100875, China}

\author{Cui-Wen Zhang}
\thanks{These authors contributed equally to this work.}
\affiliation{Key Laboratory of Beam Technology of the Ministry of Education, and School of Physics and Astronomy, Beijing Normal University, Beijing 100875, China}
\affiliation{Institute of Applied Physics and Computational Mathematics, \mbox{Beijing 100088, China}}

\author{Xue-Ren Hong}
\affiliation{\mbox{College of Physics and Electronic Engineering, Northwest Normal University, Lanzhou 730070, China}}

\author{Feng Wan}
\email{wanfeng@xjtu.edu.cn}
\affiliation{Ministry of Education Key Laboratory for Nonequilibrium Synthesis and Modulation of Condensed Matter, Shaanxi Province Key Laboratory of Quantum Information and Quantum Optoelectronic Devices, School of Physics, Xi'an Jiaotong University, Xi'an 710049, China}

\author{Jian-Xing Li}
\affiliation{Ministry of Education Key Laboratory for Nonequilibrium Synthesis and Modulation of Condensed Matter, Shaanxi Province Key Laboratory of Quantum Information and Quantum Optoelectronic Devices, School of Physics, Xi'an Jiaotong University, Xi'an 710049, China}

\author{Bai-Song Xie}
\email{bsxie@bnu.edu.cn}
\affiliation{Key Laboratory of Beam Technology of the Ministry of Education, and School of Physics and Astronomy, Beijing Normal University, Beijing 100875, China}
\affiliation{\mbox{Institute of Radiation Technology, Beijing Academy of Science and Technology, Beijing 100875, China}}

\begin{abstract}
Linearly polarized attosecond $\gamma$-ray pulses in the MeV range are generated from a cone target irradiated by a single few-cycle laser pulse. Electron layers are periodically extracted from the cone walls and subsequently accelerated. Their interaction with the counter-propagating reflected attosecond field produces high-energy photons through nonlinear Compton scattering (NCS), forming attosecond $\gamma$-ray pulses. We model this interaction using two-dimensional quantum electrodynamics particle-in-cell (QED-PIC) simulations that resolve electron spin and photon polarization during emission. The results show a shortest equivalent duration of $300\,\mathrm{as}$, with a corresponding linear polarization degree of 0.78. The photon spectrum extends to $6\,\mathrm{MeV}$, and the linear polarization degree in the high-energy range reaches 0.88. The linear polarization degree remains high when photons from both emission directions are collected over wide momentum-angle ranges. Scans over the cone opening angle and the coupled laser-plasma parameters reveal tradeoffs among photon number, mean photon energy, and polarization. Such highly polarized attosecond $\gamma$-ray pulses could be used to investigate ultrafast nuclear dynamics and polarization-dependent processes in strong-field quantum electrodynamics.
\end{abstract}

\keywords{attosecond gamma rays, photon polarization, nonlinear Compton scattering, QED-PIC simulations}

\maketitle

\section{Introduction}

Ultraintense lasers provide a laboratory route for generating high-energy $\gamma$ rays.\cite{ridgers2012,sarri2014,xue2020} Photon energy and yield are basic source characteristics, while pulse duration sets the accessible time resolution and polarization provides access to polarization-dependent observables.\cite{krausz2009,weller1992,li2020} For example, attosecond pulses can resolve electron motion in atoms and molecules.\cite{hentschel2001,morishita2007,krausz2009} Linearly polarized high-energy photons, in turn, provide angular-distribution observables in photonuclear reactions.\cite{weller1992} In nonlinear Compton scattering (NCS), their polarization can also carry signatures of the driving-laser polarization, electron spin, and interaction dynamics.\cite{li2020,tang2020,king2020,wang2024} Combining attosecond duration and high linear polarization in a high-energy photon source could therefore support studies of ultrafast processes and polarization effects in nuclear physics, laboratory astrophysics, and strong-field quantum electrodynamics.\cite{li2020,cui2025,zilges2022,utsunomiya2003}

Attosecond sources have been studied extensively in the extreme-ultraviolet and X-ray ranges.\cite{krausz2009,li2017,duris2020,heissler2012,edwards2016,zhang2020} At higher photon energies, theoretical and numerical studies of NCS and laser-plasma interactions have predicted attosecond emission at MeV energies and above.\cite{li2015,zhu2018,zhang2022cone} In parallel, electron-laser collision and laser-plasma schemes have been predicted to produce highly polarized, high-energy $\gamma$ rays.\cite{li2020,tang2020,king2020,xue2020} However, only a few studies have treated attosecond temporal structure and photon polarization together. Among these studies, a nonuniform near-critical-density plasma was predicted to generate GeV polarized attosecond $\gamma$-ray pulses with a reported duration of $760\,\mathrm{as}$ and a polarization degree of 0.60.\cite{elaji2022} In that work, polarization was derived from a separate theoretical calculation rather than evaluated during photon emission within the particle-in-cell (PIC) simulation. Another proposal used a nanofoil to generate an isolated polarized $\gamma$-ray pulse with a duration of $800\,\mathrm{as}$ and a peak linear polarization degree of 0.76.\cite{zhang2022nanofoil} This scheme requires an electron layer formed from an ultrathin foil to collide with a counterpropagating intense laser pulse under controlled timing and alignment. A third approach exploited beam instabilities to generate highly collimated polarized attosecond $\gamma$-ray pulses with \newrev{reported durations of $430$--$720\,\mathrm{as}$ for different photon-energy ranges} and an angle-resolved total linear polarization degree of 0.38, but relies on an externally preaccelerated relativistic electron beam injected into a solid-density plasma.\cite{cui2025} These studies show that attosecond duration and net photon polarization can coexist, but current approaches still face the limitations described above. More importantly, simultaneously achieving a short attosecond duration and a high net photon polarization degree remains challenging.

Our group previously proposed an attosecond $\gamma$-ray source based on a cone target irradiated by a single few-cycle laser pulse.\cite{zhang2022cone} This configuration integrates electron generation, acceleration, and scattering within the same laser-target interaction, so no externally preaccelerated electron beam or independently synchronized scattering pulse is required.\cite{zhang2022cone} The previous study established the temporal structure and spectrum of directional MeV attosecond $\gamma$-ray pulses, but did not resolve electron spin or photon polarization. It therefore remained unclear whether the \termrev{emitted attosecond $\gamma$-ray ensemble} carries a nonzero net linear polarization and whether that polarization is retained under finite momentum-angle collection.

Here, we use quantum electrodynamics particle-in-cell (QED-PIC) simulations to address these questions. The model is two-dimensional and resolves electron spin and photon polarization during emission.\cite{wan2023} The source forms photon-density structures with a shortest equivalent duration of $300\,\mathrm{as}$ and a corresponding linear polarization degree of 0.78. Compared with existing reports on \termrev{polarized attosecond $\gamma$-ray generation}, the reported values here combine a shorter pulse duration with a higher linear polarization degree. The spectrum extends to $6\,\mathrm{MeV}$, and the linear polarization degree in the high-energy range reaches 0.88. The linear polarization degree remains high when photons from both emission directions are collected over wide momentum-angle ranges. Scans over the cone opening angle and the coupled laser-plasma parameters further show tradeoffs among photon number, mean photon energy, and polarization. The energy- and angle-resolved results show how to select highly polarized, high-energy components of the attosecond $\gamma$-ray pulses.

\section{Simulation setup}

\begin{figure}[!htbp]
  \centering
\includegraphics[width=\columnwidth]{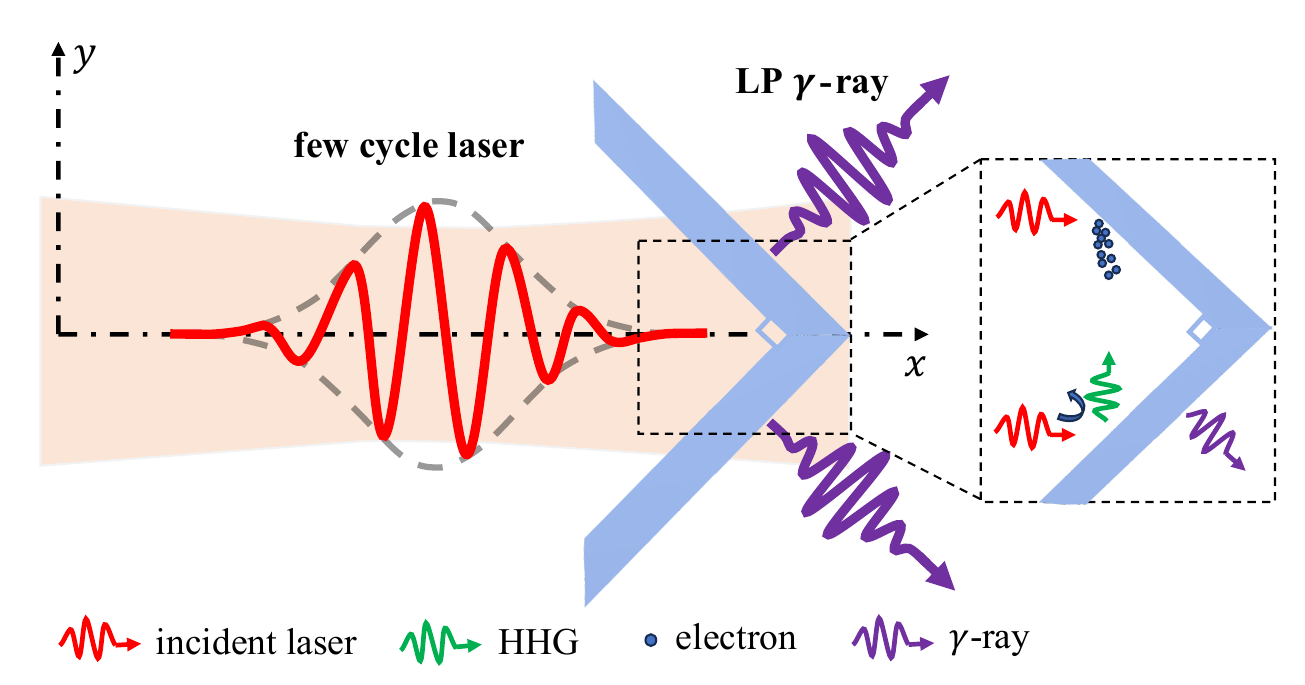}
  \caption{Generation scheme for polarized attosecond $\gamma$ rays.
A few-cycle linearly polarized laser pulse propagates along the symmetry axis
of the cone target. Oblique reflection produces a wall-normal field
that extracts electron layers. Their subsequent interaction with the
counter-propagating reflected field generates polarized $\gamma$ rays through
NCS.}
  \label{fig:fig1}
\end{figure}
Figure~\ref{fig:fig1} shows the laser--cone interaction used to generate
polarized attosecond $\gamma$ photons. A few-cycle laser pulse propagates along
the $+x$ direction, which is the symmetry axis of the cone target, and is
linearly polarized along $+y$. Although the laser is normally incident on the
target as a whole, it is obliquely incident on each inner cone wall. For the
$90^\circ$ reference configuration, the two walls therefore form symmetric
regions of oblique reflection and electron extraction. The earlier
cone-target study established the underlying three-stage process of electron
injection, acceleration, and scattering.\cite{zhang2022cone} Oblique
reflection in the relativistic oscillating mirror (ROM) regime produces
attosecond electromagnetic structures and a wall-normal field that periodically
extracts electron layers.\cite{baeva2006,heissler2012,zhang2022cone} The
charge-separation field first accelerates compressed electron bunches to tens
of MeV, while the incident and reflected laser fields modify their energy and
momentum at later stages.\cite{zhang2022cone} The electrons then emit
high-energy photons through NCS when they encounter the counter-propagating
reflected attosecond field. The present study uses this established source
mechanism and focuses on the polarization of the emitted photons.

The simulations were performed with the spin- and polarization-resolved QED-PIC
code SLIPs,\cite{wan2023} which resolves the Stokes parameters of emitted
photons and the spin state of electrons in the nonlinear QED emission process.
The photon Stokes parameters $(\xi_1,\xi_2,\xi_3)$ are defined with
respect to the transverse axes
$\hat{\mathbf P}_1\parallel\hat{\mathbf a}-\hat{\mathbf n}
(\hat{\mathbf n}\!\cdot\!\hat{\mathbf a})$ and
$\hat{\mathbf P}_2=\hat{\mathbf n}\times\hat{\mathbf P}_1$. Here,
$\hat{\mathbf n}$ and $\hat{\mathbf a}$ denote the photon-emission and
parent-electron acceleration directions, respectively.\cite{wan2023}
The reference simulation uses a two-dimensional box spanning
$18\lambda_0\times16\lambda_0$ with $1800\times1600$ cells and
80 macroparticles per cell.
The Gaussian laser pulse has a normalized amplitude
$a_0=eE_0/(m_e\omega_0c)=30$, which at
$\lambda_0=1\,\mu\mathrm{m}$ corresponds to a peak intensity
$I=1.37\times10^{18}a_0^2\simeq1.23\times10^{21}\,\mathrm{W\,cm^{-2}}$.
Here, $-e$ and $m_e$ are the electron charge and mass, $E_0$ is the electric
field amplitude, $\omega_0=2\pi/T_0$ is the laser frequency, and
$T_0=\lambda_0/c$ is the laser period. The simulated laser intensity
envelope has a full width at half maximum (FWHM) of approximately
$\tau_{\mathrm{FWHM}}=1.5T_0$, and the focal-spot radius is $w_0=3\lambda_0$.
The laser is injected through the $x_{\min}$ boundary, while the
$x_{\max}$ and both transverse boundaries use simple-outflow conditions.

The cone target consists of fully ionized carbon plasma and has an
opening angle of $90^\circ$, a wall thickness of
$d=1\lambda_0$, and an electron density of $n_e=40n_c$, where
$n_c=m_e\varepsilon_0\omega_0^2/e^2\simeq1.1\times10^{21}\,\mathrm{cm^{-3}}$
is the critical density. The initial electron ensemble is
unpolarized. These parameters give the similarity parameter
$S=n_e/(a_0n_c)=1.33$.\cite{gordienko2005} In the coupled parameter scan,
$a_0$ and $n_e$ are increased together so that $S$ remains at 1.33.
Thus, the target density is scaled in proportion to the laser amplitude, allowing
the cases to be compared under the same relativistic-similarity condition
rather than as a pure laser-intensity scan. Due to the two-dimensional geometry, the
summed photon weights are multiplied by the focal-spot radius $w_0$, taken as
the length in the third spatial dimension, to estimate the total photon
number.\cite{hou2019}

\section{Results and discussion}

Figure~\ref{fig:fig2}(a) shows the longitudinal electric-field component
$E_x/E_0$. Narrow bands of alternating positive and negative
$E_x$ follow the two inner cone surfaces, while curved $E_x$ structures are
visible outside the cone at $x>12\lambda_0$. The local amplitude reaches several
$E_0$. Although the incident field is polarized along $y$, oblique reflection
from the inclined walls produces a reflected field with a nonzero longitudinal
component $E_x$. For the $90^\circ$ cone, the wall-normal field contains this
component,
$E_\perp=E_x\cos45^\circ\pm E_y\sin45^\circ$, and periodically extracts
electron layers from the two inner surfaces.\cite{zhang2022cone}
Panel (a) therefore shows the longitudinal field structures associated
with the localized photon-density bands in panel (b).

\begin{center}
\begin{minipage}{\columnwidth}
  \centering
\includegraphics[width=\columnwidth]{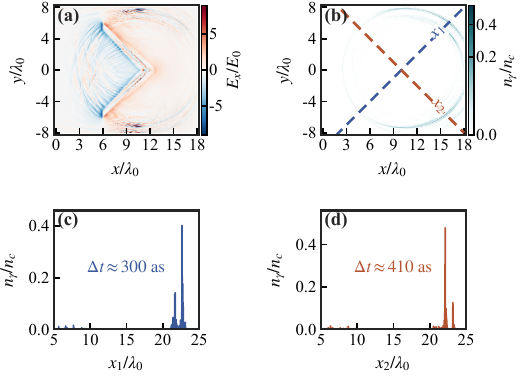}
  \captionof{figure}{Longitudinal electric-field distribution and directional temporal structure of the \termrev{$\gamma$-ray pulses}.
All panels correspond to $t=22.0T_0$ ($73.4\,\mathrm{fs}$).
(a) Longitudinal electric-field component $E_x/E_0$. (b) Photon density
$n_\gamma/n_c$ in the $x$--$y$ plane. The dashed
lines define the two spatial diagnostic directions,
$x_1$: $y=x-10\lambda_0$ and
$x_2$: $y=-x+10\lambda_0$. (c,d) Photon-density profiles along
$x_1$ and $x_2$, respectively. The labels report representative equivalent durations inferred from the FWHM of the resolved broad photon-density structures: approximately $300\,\mathrm{as}$ along $x_1$ and $410\,\mathrm{as}$ along $x_2$.}
  \label{fig:fig2}
\end{minipage}
\end{center}

Figure~\ref{fig:fig2}(b) gives the photon density. The
photon density is located mainly outside the cone on its $+x$ side and forms
curved density bands along the two diagonal directions, while the density near
the target axis is much weaker. The maximum photon density in panel (b)
is $n_\gamma/n_c\simeq0.53$. To characterize these structures, we take two
spatial lineouts through the
diagonal photon-density structures, $x_1:y=x-10\lambda_0$ and
$x_2:y=-x+10\lambda_0$. These dashed lines are spatial diagnostics rather
than momentum-angle cuts. Each lineout contains several localized density
features. We determine the representative equivalent pulse duration from
the spatial FWHM $\Delta l$ of each resolved broad density structure as
$\Delta t=\Delta l/c$, since the photons propagate at approximately $c$.

The corresponding profiles are shown in Figs.~\ref{fig:fig2}(c) and
\ref{fig:fig2}(d). For the same density-defined structures, the $x_1$ and
$x_2$ lineouts give representative equivalent durations of approximately
$300\,\mathrm{as}$ and $410\,\mathrm{as}$, with photon-weighted linear
polarization degrees of 0.78 and 0.75, respectively. The localized attosecond
photon-density structures are therefore themselves linearly polarized.
For each pulse, the Stokes parameters are averaged with photon
macroparticle weights over the photons inside its density-defined FWHM interval
and within a $\pm40\,\mathrm{nm}$ transverse tube centered on the corresponding
diagnostic line.
The $\pm45^\circ$ directions in Fig.~\ref{fig:fig2} are spatial
lineout directions for this pulse-width diagnostic. The energy- and
momentum-angle-resolved analysis below, using
$\theta_\gamma=\operatorname{atan2}(p_y,p_x)$, addresses the separate
high-energy ensemble and its angular collection.

\begin{center}
\begin{minipage}{\columnwidth}
  \centering
\makebox[\columnwidth][c]{\includegraphics[width=\columnwidth]{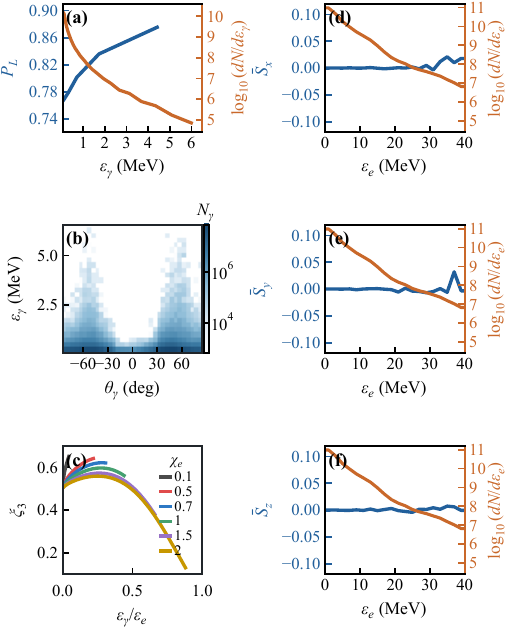}}
  \captionof{figure}{Spectral, angular, and spin-resolved properties of the \termrev{$\gamma$-ray source}.
(a) Average linear polarization degree
$P_L=\sqrt{\langle\xi_1\rangle^2+\langle\xi_3\rangle^2}$
(blue, left axis) and photon energy spectrum (orange, right axis).
(b) Photon-number distribution in the $(\varepsilon_\gamma,\theta_\gamma)$
plane. Here, $\varepsilon_\gamma$ is the photon energy and
$\theta_\gamma=\operatorname{atan2}(p_y,p_x)$, with $p_x$ and $p_y$ denoting
the $x$- and $y$-components of the photon momentum.
(c) Theoretical Stokes parameter $\xi_3$ as a function of
$\varepsilon_\gamma/\varepsilon_e$ for different values of $\chi_e$.
(d--f) Electron energy spectra (orange, right axes) and average spin components
$\overline{S}_x$, $\overline{S}_y$, and $\overline{S}_z$
(blue, left axes), respectively.}
  \label{fig:fig3}
\end{minipage}
\end{center}

Figure~\ref{fig:fig3}(a) presents the photon energy spectrum (orange) and the
linear polarization degree (blue). The photon yield is concentrated below
$1\,\mathrm{MeV}$, and the spectrum decreases steeply to a cutoff energy of
approximately $6\,\mathrm{MeV}$. By contrast, $P_L$ increases from
0.77 in the lowest energy bin ($0.05$--$0.2\,\mathrm{MeV}$) to
0.88 for photons in the $3$--$6.5\,\mathrm{MeV}$ range. As a result, selecting the high-energy photons
provides a more strongly polarized but less numerous photon sample. The final blue
point at $4.39\,\mathrm{MeV}$ represents the full $3$--$6.5\,\mathrm{MeV}$ range;
it does not indicate that the polarization vanishes above $4.39\,\mathrm{MeV}$.

Figure~\ref{fig:fig3}(b) shows the energy-resolved angular distribution of the
emitted photons. For $\varepsilon_\gamma\geq3\,\mathrm{MeV}$, the photon number is
highest near $\theta_\gamma=-52.5^\circ$ and $+62.5^\circ$. Integrating these
high-energy photons over energy gives an angular full width at half maximum of
about $10^\circ$ for each emission branch. This concentration into two narrow
branches becomes weaker at lower energies: the lower-energy photons cover a
wider angular range, and the photon density along the target axis
($\theta_\gamma=0$) remains much lower than at the two off-axis maxima.
The two branches follow the motion of the radiating electrons. An
ultrarelativistic electron emits radiation through NCS within a narrow angle
around its instantaneous momentum. Electrons extracted from the two cone walls
acquire transverse momenta of opposite sign while they are accelerated forward,
producing the photon maxima at negative and positive
$\theta_\gamma$.\cite{zhang2022cone}

To explain why $P_L$ increases with photon energy, we first write the linear
polarization degree in terms of the two linear Stokes components,
$P_L=\sqrt{\langle\xi_1\rangle^2+\langle\xi_3\rangle^2}$. In every photon-energy
bin, $|\langle\xi_1\rangle|<6.6\times10^{-3}$, whereas
$\langle\xi_3\rangle$ increases from 0.77 to approximately 0.88. Because
$|\langle\xi_1\rangle|\ll|\langle\xi_3\rangle|$,
$P_L\simeq|\langle\xi_3\rangle|$. We therefore analyze $\xi_3$ to explain the
increase of $P_L$ with photon energy. Xue \textit{et al.} showed that the
single-emission $\xi_3$ first increases slightly and then decreases with
$r=\varepsilon_\gamma/\varepsilon_e$ in the stronger-recoil regime.\cite{xue2020}
This result concerns $\xi_3$ at specified $\chi_e$ and $r$, whereas
Fig.~\ref{fig:fig3}(a) gives $P_L$ for photons grouped by absolute
energy.

To determine whether a net electron spin contributes to this trend, panels
(d--f) show the electron energy spectrum together with the mean spin components.
Only the $0$--$40\,\mathrm{MeV}$ range is displayed because it contains most
of the electrons; the spectral density has
already decreased by more than three orders of
magnitude at $30\,\mathrm{MeV}$. Below this energy, the three mean spin
components in each energy interval satisfy
$|\overline S_{x,y,z}|<7\times10^{-3}$. Above
$30\,\mathrm{MeV}$, few electrons remain and the mean spin components
fluctuate in sign, so this region does not establish a common net spin
direction. Most
electrons lie below
$30\,\mathrm{MeV}$, where the ensemble has no appreciable net spin polarization.
This does not mean that individual electrons have zero spin.
The energy-resolved averages do not imply that spin-dependent terms vanish in
every emission event. We therefore use the
spin-averaged rate only as a reference for interpreting the overall trend; the
QED-PIC calculation still includes the spin dependence of individual emissions.

The spin- and polarization-resolved photon-emission rate in NCS can be
written as\cite{li2020,xue2020,wan2023}
\begin{equation}
 \frac{d^2W_{\mathrm{rad}}}{du\,dt}
 =\frac{W_R}{2}\left(F_0+\xi_1F_1+\xi_2F_2+\xi_3F_3\right),
 \label{eq:general-rate}
\end{equation}
where $W_R$ is a common positive prefactor.
Because the initial electron ensemble is unpolarized and the final
electron spin is not observed, we average over the two initial spin states and
sum over the two final spin states. After these operations, all
electron-spin-dependent contributions cancel, so $F_1=F_2=0$, whereas $F_0$
and $F_3$ reduce to the spin-independent functions below (the full
spin-resolved coefficients are given in Appendix~\ref{app:spin-coefficients}):
\begin{align}
 F_0={}&-(2+u)^2\left[\mathcal{I}_{1/3}(u')-2K_{2/3}(u')\right] \nonumber\\
 &+u^2\left[\mathcal{I}_{1/3}(u')+2K_{2/3}(u')\right], \nonumber\\
 F_3={}&4(1+u)K_{2/3}(u'),
 \qquad
 \mathcal{I}_{1/3}(u')=\int_{u'}^\infty K_{1/3}(z)\,dz,
 \label{eq:f0f3}
\end{align}
where $u=r/(1-r)$, $u'=2u/(3\chi_e)$, and
$r=\varepsilon_\gamma/\varepsilon_e$, with $\varepsilon_e$ denoting the electron
energy immediately before emission. Here, $K_\nu$ is the modified Bessel function
of the second kind, and $\chi_e$ measures the field in the electron rest frame and
hence the importance of quantum recoil. The Stokes vector of the mixed photon
state is $\bm\xi^{(\mathrm{mix})}=(F_1,F_2,F_3)/F_0$ in this
formulation.\cite{li2020,wan2023} The spin-averaged result therefore gives
$\xi_1=\xi_2=0$ and $\xi_3=F_3/F_0$, consistent with the small
simulated values $|\langle\xi_1\rangle|<6.6\times10^{-3}$ in the same photon
energy intervals.

Figure~\ref{fig:fig3}(c) plots the theoretical Stokes parameter $\xi_3$ against
the photon energy fraction $r=\varepsilon_\gamma/\varepsilon_e$. At
$\chi_e=0.1$ and 0.5, $\xi_3$ increases throughout the displayed range. Thus,
$\xi_3$ is larger at larger photon-to-electron energy ratios in this parameter
range. At $\chi_e=0.7$, the curve reaches a
shallow maximum near $r=0.28$. For $\chi_e=1$, 1.5, and 2, the maximum occurs at
$r\simeq0.25$--0.28 and is followed by a clear decrease. The spin-averaged rate
therefore contains both behaviors: an increase at small $\chi_e$ and a maximum
followed by a decrease when quantum recoil is stronger. This change of regime
is consistent with the turnover reported by Xue \textit{et al.} in the
stronger-recoil regime.\cite{xue2020} Panel (c) therefore provides a qualitative
reference for the energy dependence of $P_L$ in panel (a), rather than
a point-by-point prediction. The values in panel (a) are calculated directly
from the QED-PIC photons in each energy interval.

\begin{figure}[!t]
  \centering
\includegraphics[width=0.92\columnwidth]{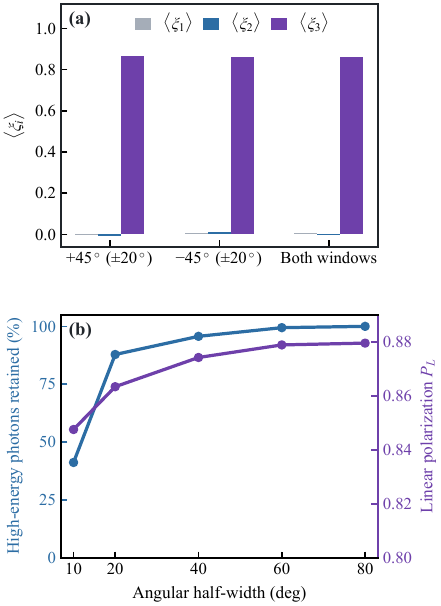}
  \caption{Momentum-angle collection of polarized high-energy photons centered on the $\pm45^\circ$ reference axes.
(a) Average Stokes parameters $\langle\xi_1\rangle$, $\langle\xi_2\rangle$, and
$\langle\xi_3\rangle$ for photons with $\varepsilon_\gamma\geq3\,\mathrm{MeV}$
selected within two angular ranges centered at momentum angles
$\theta_\gamma=+45^\circ$ and $-45^\circ$, each with an angular half-width of
$20^\circ$. The results for
each angular range and for their union are shown. (b) Fraction of the
$\varepsilon_\gamma\geq3\,\mathrm{MeV}$ photons retained (blue) and their linear
polarization degree $P_L$ (purple) as the angular half-width of both ranges is
varied. The $\pm45^\circ$
axes correspond to the spatial pulse directions used in Fig.~2; the angular
selection itself is made in momentum space.
All Stokes averages are photon-weighted after transformation to the
common laboratory-frame convention based on the projected $y$ axis.}
  \label{fig:fig4}
\end{figure}

Figure~\ref{fig:fig4}(a) shows the Stokes parameters for the two high-energy
branches and for their combined collection. For
photons with $\varepsilon_\gamma\geq3\,\mathrm{MeV}$, the
$+45^\circ(\pm20^\circ)$ and $-45^\circ(\pm20^\circ)$ angular ranges have
nearly identical positive $\langle\xi_3\rangle$, 0.87 and 0.86,
respectively.
Their $\langle\xi_1\rangle$ and $\langle\xi_2\rangle$ values remain close to
zero. This further supports the treatment used in the energy-resolved analysis
in Fig.~\ref{fig:fig3}(a), where $\langle\xi_1\rangle$ was small and $P_L$ was
mainly determined by $\langle\xi_3\rangle$. When the two angular ranges are combined,
$\langle\xi_3\rangle=0.86$ and $P_L=0.86$. The two angular ranges therefore
have the same dominant linear-polarization component in the
common laboratory basis,
so their combination does not average to a weakly polarized sample.
Before combining photons with different propagation directions, we
express each Stokes vector in a laboratory-anchored transverse basis. Its first
axis is
$\hat{\mathbf e}_{y\perp}=[\hat{\mathbf y}
-\hat{\mathbf n}(\hat{\mathbf n}\!\cdot\!\hat{\mathbf y})]/
\sqrt{1-(\hat{\mathbf n}\!\cdot\!\hat{\mathbf y})^2}$, the projection of the
laboratory $y$ axis onto the plane normal to the photon direction
$\hat{\mathbf n}$. The linear components $(\xi_1,\xi_3)$ are rotated into this
basis by the standard $2\psi$ transformation before photon-weighted averaging.
In the present two-dimensional data, every stored $\hat{\mathbf P}_1$ axis is
already parallel or antiparallel to $\hat{\mathbf e}_{y\perp}$, so the
averages are unchanged.\cite{wan2023}

Figure~\ref{fig:fig4}(b) shows the retained photon fraction and $P_L$ for
different angular half-widths of the two collection ranges. With a $5^\circ$
angular half-width, the two ranges retain 20.8\% of the photons above
$3\,\mathrm{MeV}$ and give
$P_L=0.82$. A $20^\circ$ half-width raises the retained fraction to
87.8\% while keeping $P_L=0.86$. At $40^\circ$, the retained fraction is
95.7\% and $P_L=0.87$. The increasing retained fraction follows from the wider angular
selection. The useful point is that including high-energy photons farther from
the reference axes does not lower the linear polarization degree. Most
high-energy photons can therefore be collected over a moderately wide angular
range while maintaining a high linear polarization degree.
It is worth noting that the $\pm45^\circ$ axes are used only as the centres of
the collection windows. They are tied to the spatial pulse directions in Fig.~2,
where the photon density was sampled along two diagonal lineouts. The angular
selection in Fig.~\ref{fig:fig4} is based instead on the photon momentum angle
because the photon momentum fixes its propagation direction and therefore
determines whether it enters an angular collection window. After the main-case emission and collection properties have been established, the same diagnostics can be used to examine how the source changes with target geometry and laser-density scaling.

\begin{figure}[!t]
  \centering
\makebox[\columnwidth][c]{\includegraphics[width=1.06\columnwidth]{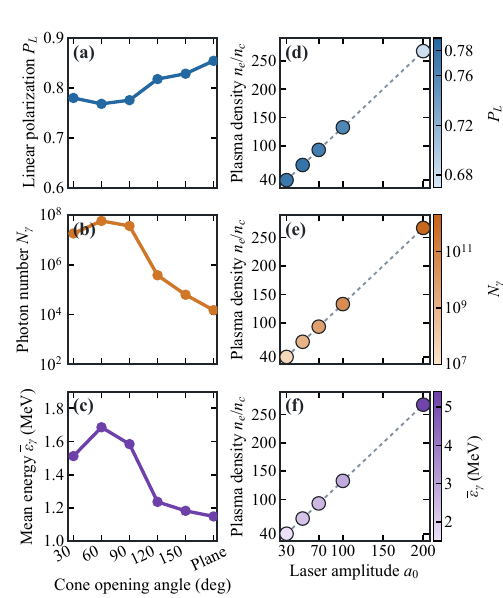}}
  \caption{Parameter dependence of photon polarization, number, and mean energy.
(a--c) Linear polarization degree $P_L$, photon number $N_\gamma$, and mean
photon energy $\overline{\varepsilon}_\gamma$ as functions of cone opening
angle for $a_0=30$ and $n_e=40n_c$. The label ``Plane'' denotes the
$180^\circ$ limit. (d--f) The same quantities along the coupled
$a_0$--$n_e$ scan with $S=n_e/(a_0n_c)=1.33$. Colour indicates the value at
each simulated point. The dotted lines connect neighboring simulated points;
no fitting is performed. All results
are taken at $t=22.0T_0$ ($73.4\,\mathrm{fs}$). $N_\gamma$ and
$\overline{\varepsilon}_\gamma$ are calculated for photons with
$\varepsilon_\gamma\geq1\,\mathrm{MeV}$. The $P_L$ values use the full
photon spectrum.}
  \label{fig:fig5}
\end{figure}

Figure~\ref{fig:fig5}(a--c) shows how the source properties change with the
cone opening angle at fixed $a_0=30$ and $n_e=40n_c$. The linear polarization
degree remains between 0.77 and 0.78 for $30^\circ$--$90^\circ$ and then increases to
0.82, 0.83, and 0.85 for $120^\circ$, $150^\circ$, and the
plane-target limit, respectively. This increase, however, is accompanied by a strong
reduction in the photon output above $1\,\mathrm{MeV}$. The photon number is
largest at $60^\circ$ ($5.7\times10^7$), remains of order $10^7$ at
$30^\circ$ and $90^\circ$, and then falls to $3.8\times10^5$ at $120^\circ$,
$6.2\times10^4$ at $150^\circ$, and about $1.5\times10^4$ in the plane-target
limit. The mean photon energy follows the same loss of high-energy output:
it is $1.51$--$1.69\,\mathrm{MeV}$ for $30^\circ$--$90^\circ$ and decreases to
$1.24\,\mathrm{MeV}$ at $120^\circ$, $1.18\,\mathrm{MeV}$ at $150^\circ$, and
$1.15\,\mathrm{MeV}$ for the plane target. This decrease follows the change in
the collision geometry. In this laser--cone interaction, high-energy photons
are produced when escaped electrons interact nearly head-on with the
counter-propagating reflected attosecond electromagnetic field.\cite{zhang2022cone} Increasing the cone
opening beyond $90^\circ$ weakens this counter-propagating interaction and the
cone focusing, so
fewer electrons contribute to high-energy NCS emission. The
plane-target limit removes the two-wall cone geometry and leaves
only a much weaker high-energy photon population. Thus, the higher polarization
in the larger-opening-angle and plane-target cases does not by itself indicate
a better high-energy photon source.

Figures~\ref{fig:fig5}(d--f) show the fixed-$S$ scan defined in the
simulation setup. In this scan, increasing $a_0$ increases the field amplitude
and electron energy gain before emission, so the emitted photons can reach
higher energies in NCS. At the same time, the
higher $n_e$ supplies more radiating
electrons while the relative density is scaled to the laser amplitude, which
raises the number of photons above $1\,\mathrm{MeV}$. The photon number above
$1\,\mathrm{MeV}$ increases from $3.6\times10^7$ at $a_0=30$ to
$1.4\times10^{12}$ at $a_0=200$, and the mean photon energy rises from
$1.58$ to $5.28\,\mathrm{MeV}$. The polarization behaves differently:
$P_L$ remains between 0.77 and 0.78 for $a_0=30$--70, decreases to 0.76 at
$a_0=100$, and reaches 0.68 at $a_0=200$. This behavior is consistent with
the energy-resolved analysis in Fig.~3: increasing the field strength and
density enhances photon production and photon energy, whereas stronger recoil
conditions are expected to make it harder to preserve the same linear
polarization degree.
The fixed-$S$ scan therefore identifies a second trade-off, between higher
yield and photon energy on one side and polarization retention on the other.
Because $a_0$ and $n_e$ are varied together, these data do not isolate their
individual contributions.

\section{Conclusion}

\concrev{In summary, we use spin- and polarization-resolved QED-PIC simulations
to determine the polarization properties of attosecond $\gamma$-ray pulses
generated from a cone target irradiated by a single few-cycle laser pulse.}
Oblique reflection produces a wall-normal field that extracts surface
electrons. After acceleration, the electrons encounter the counter-propagating
reflected attosecond field and emit high-energy photons through NCS, forming
attosecond $\gamma$-ray pulses. The results
show that the laser--cone interaction produces linearly polarized attosecond
$\gamma$-ray pulses. The shortest
equivalent duration is $300\,\mathrm{as}$, with a corresponding
linear polarization degree of 0.78. The photon spectrum extends to
$6\,\mathrm{MeV}$, and the linear polarization degree reaches 0.88 at the
high-energy end.
\newrev{Compared with existing reports on \termrev{polarized attosecond $\gamma$-ray generation},
this result provides a shorter reported duration while retaining high
pulse-resolved linear polarization.}\cite{elaji2022,zhang2022nanofoil,cui2025}
Taking the spatial pulse directions at $\pm45^\circ$ as reference axes,
the linear polarization degree remains high when photons from both
emission directions are collected over wider momentum-angle ranges. A
relatively wide angular collection range can therefore be used.

We also examined the effects of the cone opening angle and the coupled
$a_0$--$n_e$ variation at fixed $S=n_e/(a_0n_c)$. The linear polarization
degree changes little for opening angles from $30^\circ$ to $90^\circ$ and
increases at larger opening angles, but the photon number and mean energy then
decrease strongly. Along the fixed-$S$ path, increasing $a_0$ and $n_e$
increases the photon number and mean energy but reduces the linear polarization
degree. The target and laser--plasma parameters must therefore be chosen
according to the required balance among photon number, energy, and
polarization. This combination of attosecond duration, MeV photon energy, high
linear polarization, and finite-angle collection provides a source concept for
ultrafast, polarization-resolved photonuclear measurements.

\begin{acknowledgments}
\ackrev{This work was supported by the National Natural Science Foundation
of China (Grant Nos.\ 12375240, 12475249, 12535015, and 12565023) and the
National Key Research and Development (R\&D) Program
(Grant No.\ 2024YFA1612700).}
The computations were performed using computing resources at Xi'an
Jiaotong University and Beijing Normal University.
\end{acknowledgments}

\appendix
\section{Spin-resolved NCS coefficients}
\label{app:spin-coefficients}

For reference, the full spin-resolved coefficients entering
Eq.~\eqref{eq:general-rate} are listed below. They are written in the locally
constant field approximation and follow Refs.~\onlinecite{li2020,xue2020,wan2023}.
Define $\bm b=\hat{\bm n}\times\hat{\bm a}$,
$S_{if}=\bm S_i\cdot\bm S_f$, and
$\mathcal I_{1/3}(u')=\int_{u'}^\infty K_{1/3}(z)\,dz$.

Here, $\bm S_i$ and $\bm S_f$ are the initial- and final-electron spin
polarization vectors, respectively, with $|\bm S_i|=|\bm S_f|=1$ for the
resolved spin states; $\hat{\bm n}$ is the photon-emission direction; and
$\hat{\bm a}$ is the unit direction of the parent-electron acceleration. The quantities $u$,
$u'$, $r$, $\chi_e$, and $K_\nu$ are defined below Eq.~\eqref{eq:f0f3}.

\begin{align}
F_0={}&-(2+u)^2\left[\mathcal I_{1/3}(u')-2K_{2/3}(u')\right]
\nonumber\\
&\quad\times(1+S_{if})
\nonumber\\
&+u^2(1-S_{if})\left[\mathcal I_{1/3}(u')+2K_{2/3}(u')\right]
\nonumber\\
&+2u^2S_{if}\mathcal I_{1/3}(u')
\nonumber\\
&-(4u+2u^2)(\bm S_f+\bm S_i)\cdot\bm b\,K_{1/3}(u')
\nonumber\\
&-2u^2(\bm S_f-\bm S_i)\cdot\bm b\,K_{1/3}(u')
\nonumber\\
&-4u^2\left[\mathcal I_{1/3}(u')-K_{2/3}(u')\right]
\nonumber\\
&\quad\times(\bm S_i\cdot\hat{\bm n})(\bm S_f\cdot\hat{\bm n}),
\label{eq:app-f0}
\end{align}

\begin{widetext}
\begin{align}
F_1={}&-2u^2\mathcal I_{1/3}(u')
\left[(\bm S_i\cdot\hat{\bm a})(\bm S_f\cdot\bm b)
+(\bm S_f\cdot\hat{\bm a})(\bm S_i\cdot\bm b)\right] \nonumber\\
&+4u\left[(1+u)(\bm S_i\cdot\hat{\bm a})
+\bm S_f\cdot\hat{\bm a}\right]K_{1/3}(u')
+2u(2+u)\hat{\bm n}\cdot(\bm S_f\times\bm S_i)K_{2/3}(u'),
\label{eq:app-f1}\\[2pt]
F_2={}&-\Big\{2u^2\left[(\bm S_i\cdot\hat{\bm n})(\bm S_f\cdot\bm b)
+(\bm S_f\cdot\hat{\bm n})(\bm S_i\cdot\bm b)\right]
+2u(2+u)\hat{\bm a}\cdot(\bm S_f\times\bm S_i)\Big\}K_{1/3}(u') \nonumber\\
&-4u\left[\bm S_i\cdot\hat{\bm n}
+(1+u)\bm S_f\cdot\hat{\bm n}\right]\mathcal I_{1/3}(u')
+4u(2+u)(\bm S_i+\bm S_f)\cdot\hat{\bm n}\,K_{2/3}(u'),
\label{eq:app-f2}\\[2pt]
F_3={}&4\left[1+u+\left(1+u+\frac{u^2}{2}\right)S_{if}
-\frac{u^2}{2}(\bm S_i\cdot\hat{\bm n})(\bm S_f\cdot\hat{\bm n})\right]K_{2/3}(u') \nonumber\\
&+2u^2\left[(\bm S_i\cdot\bm b)(\bm S_f\cdot\bm b)
-(\bm S_i\cdot\hat{\bm a})(\bm S_f\cdot\hat{\bm a})\right]\mathcal I_{1/3}(u') \nonumber\\
&-4u\left[(1+u)\bm S_i\cdot\bm b+\bm S_f\cdot\bm b\right]K_{1/3}(u').
\label{eq:app-f3}
\end{align}
\end{widetext}

\bibliography{Zhang_et_al_Polarized_Attosecond_Gamma_Ray_Emission}

\end{document}